\documentclass[preprin,showpacs,preprintnumbers,eqsecnum,amsmath,amssymb,twoside]{revtex4}
\usepackage{graphics,epsfig}
\usepackage{graphicx}
\usepackage{dcolumn}
\usepackage{bm}

\begin{document}
\baselineskip=0.5 cm
\title{{\bf Gravitational quasinormal modes of black holes in Einstein-aether theory}}

\author{Chikun Ding }
\thanks{ Email: dingchikun@163.com; Chikun\_Ding@huhst.edu.cn }

  \affiliation{Department of Physics, Hunan University of Humanities, Science and Technology, Loudi, Hunan
417000, P. R. China\\
Key Laboratory of Low Dimensional
Quantum Structures and Quantum Control of Ministry of Education,
and Synergetic Innovation Center for Quantum Effects and Applications,
Hunan Normal University, Changsha, Hunan 410081, P. R. China}

\vspace*{0.2cm}
\begin{abstract}
\baselineskip=0.5 cm
\begin{center}
{\bf Abstract}
\end{center}

The local Lorentz violation (LV) in gravity sector should show itself in derivation of the characteristic quasinormal modes (QNMs) of black hole mergers from their general relativity case. In this paper, I study QNMs  of the gravitational field perturbations to Einstein-aether black holes and, at first compare them to those in Schwarzschild black hole, and then some other known LV gravity theories. By comparing to Schwarzschild black hole, the first  kind aether black holes have larger damping rate and the second ones have lower damping rate. And they all have smaller real oscillation frequency of QNMs. By comparing to some other LV theories, the QNMs of the first kind aether black hole are similar to that of the QED-extension limit of standard model extension, non-minimal coupling to Einstein's tensor and massive gravity theories.  While as to the second kind aether black hole, they are similar to those of the noncommutative gravity theories and Einstein-Born-Infeld theories. These similarities may imply that LV in gravity sector and LV in matter sector have some intrinsic connections.

\end{abstract}

\pacs{04.50.Kd, 04.70.Dy, 04.30.-w} \maketitle

\vspace*{0.2cm}

\section{Introduction}
LIGO (Laser Interferometer
Gravitational wave Observatory) has detected gravitational wave(GW) from a binary black hole coalescence for the fourth time, on August 14, 2017 (GW170814) \cite{abbott2017}. It provides a direct confirmation for the existence of a black hole and, confirms that black hole mergers are common in the universe. The black-hole binary systems are intrinsically strongly gravitating objects that curve spacetime dramatically, and the detections of GW from them give us opportunity and an ideal tool to stress test general relativity (GR) \cite{abbott2}. Some of the detections are used to test alternative theories of gravity where Lorentz invariance (LI) is broken which affects the dispersion relation for GW \cite{driggers}. For the first time, they used GW170104 to put upper limits on the magnitude of Lorentz violation (LV) tolerated by their data and found that the bounds are important. Sotiriou  \cite{sotiriou} argued that when higher order corrections to dispersion relation of GW are present, there will be a scalar excitation which travels at a speed different from that of the standard GW polarizations or light. Hence, a smoking-gun observation of LV would be the direct detection of this scalar wave.

Lorentz invariance(LI) is one of the fundamental principles of GR and the standard model(SM) of particles and fields. Why consider LV? Because LI may not be an exact symmetry at all energies  \cite{mattingly}, particularly when one considering quantum gravity effect, it should not be applicable. Though both GR and SM based on LI and the background of spacetime, they handle their entities in profoundly different manners. GR is a classical field theory in curved spacetime that neglects all quantum properties of particles; SM is a quantum field theory in flat spacetime that neglects all gravitational effects of particles. For collisions of particles of $10^{30}$ eV energy (energy higher than Planck scale), the gravitational interactions predicted by GR are very strong and gravity should not be negligible\cite{camelia}. So in this very high energy scale,  one have to consider merging SM with GR in a single unified theory, known as "quantum gravity", which remains a challenging task. Lorentz symmetry is a continuous spacetime symmetry and cannot exist in a discrete spacetime. Therefore quantization of spacetime at energies beyond the Planck energy, Lorentz symmetry is invalid and one should reconsider giving up LI. There are some phenomena of LV. On the SM side, there is an {\it a priori }unknown physics at high-energy scales that could lead to a spontaneous breaking of LI by giving an expectation value to certain non-SM fields that carry Lorentz indices\cite{bolokhov}. LI also leads to divergences in quantum field theory which can be cured with a short distance of cutoff that breaks it \cite{jacobson}. On the GR side, astrophysical observations suggest that the high-energy cosmic rays above the Greisen-Zatsepin-Kuzmin cutoff are a result of LV\cite{carroll}.

 Thus, the study of LV is a valuable tool to probe the foundations of modern physics. These studies include LV in the neutrino sector \cite{dai}, the standard-model extension \cite{colladay}, LV in the non-gravity sector \cite{coleman}, and LV effect on the formation of atmospheric showers \cite{rubtsov}.
Einstein-aether theory can be considered as an effective description of Lorentz symmetry breaking in the gravity sector and has been extensively used in order to obtain quantitative constraints on Lorentz-violating gravity\cite{jacobson2}.

In Einstein-aether theory, the background tensor fields $u^a$ break the Lorentz symmetry only down to a rotation subgroup by the existence of a preferred time direction at every point of spacetime. The introduction of the aether vector allows for some novel effects, e.g., matter fields can travel faster than the speed of light \cite{jacobson3}, dubbed superluminal particle. Due the existence of the superluminal gravitational modes, so the corresponding light-cones can be completely flat, and the causality is more like that of Newtonian theory\cite{greenwald}. It is the
 universal horizons that can trap excitations traveling at arbitrarily high velocities.
 Recently, two exact charged black hole solutions and their Smarr formula on universal horizons in 4- and 3-dimensional spacetime were found by Ding {\it et al} \cite{ding,ding2}. Constraints on Einstein-aether theory were studied by Oost {\it et al} \cite{oost} and, gravitational wave studied by Gong {\it et al} \cite{gong2018} after GW170817. Other studies on universal horizons can be found in \cite{tian}.

In Ref. \cite{ding2016}, Ding {\it et al} studied Hawking radiation from the charged Einstein aether black hole and found that i) the universal horizon seems to be no role on the process of radiating luminal or subluminal particles and, its temperature is dependant on $z$, $T_{UH}=(z-1)\kappa_{uh}/z\pi$, where $z$ characterizes the species of the particles; while ii) the Killing horizon seems to be no role on superluminal particle radiation. Since up to date, the particles with speed higher than vacuum light speed aren't yet found, we here consider only subluminal or luminal particles perturbation to these LV black holes. In 2007, Konoplya {\it et al} \cite{konoplya} studied the gravitational perturbations of the non-reduced Einstein aether black holes and found that both the real part and the absolute imaginary part of QNMs increase with the aether coefficient $c_1$.

Recently, Ding studied the scalar and electromagnetic perturbations of the first and second Einstein-aether black holes, and found \cite{ding2017} that their characteristics are similar to that of another LV model---the QED(quantum electrodynamics)-extension  limit of SME (SM extension) \footnotemark\footnotetext{For SME, see Appendix in the reference \cite{ding2017}}. What is about their gravitational perturbations? There are two types of perturbations of a black hole: adding fields to the black hole spacetime or perturbing the black hole metric itself. Then the scalar and electromagnetic perturbations are of the first type. After the coalescence of a binary black holes or the formation of a black hole by collapse, the black hole is in a perturbed state, \begin{equation}g_{\mu\nu}=g_{\mu\nu}^0+\delta g_{\mu\nu},\end{equation}where the metric $g_{\mu\nu}^0$ is of the nonperturbed black hole when all perturbations have been damped.  This is the second type --- gravitational perturbation which is important for emitting gravitational waves. In the linear approximation, the perturbations $\delta g_{\mu\nu}$ are supposed to be much less than the background $\delta g_{\mu\nu}\ll g_{\mu\nu}^0$. The background $g_{\mu\nu}^0$ can be Schwarzschild or Einstein-aether black hole solutions.

In this paper, I study the gravitational QNMs for two kinds of Einstein-aether black holes and compare them to Schwarzschild black hole to find some derivations. And I also compare them to black holes in some LV theories to find some connections between these theories.  The plan of rest of our paper is organized as follows. In Sec. II, I review briefly the Einstein aether black holes and the sixth order WKB(Wentzel-Kramers-Brillouin) method. In Sec. III, I adopt to
the sixth order WKB method and obtain the perturbation frequencies of the first kind Einstein aether black holes. In Sec. IV, I discuss the QNMs for the second kind Einstein aether black hole. In Sec. V, I present a summary. In Appendix, I briefly introduce some Lorentz violating theories, i.e., nonminimal coupling, massive gravity, noncommutative and Einstein-Born-Infeld theories.

\section{Einstein aether black holes and WKB method}
The general action for the Einstein-aether theory can be constructed by assuming that: (1) it is general covariant; and (2)  it  is a function of only the spacetime metric $g_{ab}$ and a unit timelike vector $u^a$, and involves  no more than two derivatives of them, so that the resulting field equations are second-order differential equations of  $g_{ab}$ and   $u^a$. Then,  the  Einstein aether theory to be studied   in this paper is
 described by the  action \cite{garfinkle},
\begin{eqnarray}
\mathcal{S}=
\int d^4x\sqrt{-g}\Big[\frac{1}{16\pi G_{\ae}}(\mathcal{R}+\mathcal{L}_{\ae})\Big], \label{action}
\end{eqnarray}
where $G_{\ae}$ is the aether gravitational constant, $\mathcal{L}_{\ae}$ is the aether Lagrangian density  \begin{eqnarray}
-\mathcal{L}_{\ae}=Z^{ab}_{~~cd}(\nabla_au^c)(\nabla_bu^d)-\lambda(u^2+1)
\end{eqnarray}
with
\begin{eqnarray}
Z^{ab}_{~~cd}=c_1g^{ab}g_{cd}+c_2\delta^a_{~c}\delta^b_{~d}
+c_3\delta^a_{~d}\delta^b_{~c}-c_4u^au^bg_{cd}\,,
\end{eqnarray}
where $c_i (i = 1, 2, 3, 4)$ are coupling constants of the theory.
The aether Lagrangian density is therefore the sum of all possible terms for the aether field $u^a$ up to mass dimension two, and the constraint term $\lambda(u^2 + 1)$ with the Lagrange multiplier $\lambda$ implementing the normalization condition $u^2=-1$.
The equations of motion, obtained by varying the action (\ref{action}) with respect to $g_{ab},u^a,\lambda$, can be found in Ref. \cite{ding,garfinkle}.

There are a number of theoretical and observational bounds on the coupling constants $c_i$ \cite{jacobson2,oost,yagi,jacobson5}, e.g., requiring stability and absence of gravitational Cherenkov radiation for theoretical constraint, Solar System and cosmological observations, etc. The strong field such as binary pulsar gives more stringent constraints on the couplings, $c_{13}\lesssim0.03$, where $c_{13}\equiv c_1+c_3$, and so on. But this analysis of constraints is only valid in the small coupling region $c_i\ll1$. Under the condition that sufficiently large sensitivity and large couplings, the Einstein-aether terms can in principle dominate over the GR ones \cite{yagi}. Recently, Oost {\it et al} find  that the GW170817 and GRB170817 events provides much more severe constraint that $|c_{13}|<10^{-15}$ \cite{oost}. However, I here don't concern these severe constraints and would draw our attention on the LV gravity theory itself for theoretical interest.
Therefore from this point of view,  I impose the following
theoretical constraints \cite{berglund2012},
\begin{equation}
\label{CDs}
0\leq c_{14}<2,\quad 2+c_{13}+3c_2>0,\quad 0\leq c_{13}<1.
\end{equation}

The static, spherically symmetric metric for Einstein-aether black hole spacetime can
be written in the form
\begin{eqnarray}
ds^2 = -f(r)\,dt^2 + \frac{dr^2}{f(r)} + r^2(d\theta^2
+\sin^2\theta d\phi^2)\,. \label{metric}
\end{eqnarray}
There are two kinds of exact solutions \cite{berglund2012,ding}. In the first case $c_{14}=0,\;c_{123}\neq0$ (termed the first kind aether black hole), the metric function is
\begin{eqnarray}
\label{sol1} f(r)=1-\frac{2M}{r}-I\Big(\frac{2M}{r}\Big)^4,\;\;I=\frac{27c_{13}}{256(1-c_{13})}.
\end{eqnarray}
If the coefficient $c_{13}=0$, then it reduces to Schwarzschild black hole. The quantity $M$ is the mass of the black hole spacetime. In this case, the aether field has no contribution to black hole mass.
In the second case $c_{14}\neq0,\;c_{123}=0$ (termed the second kind aether black hole), the metric function is
\begin{eqnarray}
\label{sol2} f(r)=1-\frac{2M}{r}-J\Big(\frac{M}{r}\Big)^2,\;\;J=\frac{c_{13}-c_{14}/2}{1-c_{13}}.
\end{eqnarray}
Here $c_{13}\geq c_{14}/2$ (cf. 4.26 in \cite{ding}).
In this case, the aether field contributes spacetime mass as $M_{\ae}=-c_{14}M_{ADM}/2$ \cite{ding}, where $M_{ADM}$ is Komar mass. If the coefficient $c_{13}=c_{14}/2$, it also reduces to Schwarzschild black hole.

To gravitational field perturbations, we shall neglect small perturbations of aether field, keeping only linear perturbations of Ricci tensor for simplicity. According to Chandrasekhar designations \cite{chand}, the general form of the perturbed metric is
\begin{eqnarray}
ds^2=e^{2\nu}dt^2-e^{2\psi}(d\phi-\sigma dt-q_rdr-q_\theta d\theta)^2-e^{-2\mu_2}dr^2-e^{-2\mu_3}d\theta^2,
\end{eqnarray}
where $e^{2\nu}=e^{2\mu_2}=f(r)$, $e^{2\mu_3}=r^2, e^{2\psi}=r^2\sin^2\theta$ and $\sigma=q_r=q_\theta=0$ for non-perturbed case. The perturbations will lead to non-vanishing values of $\sigma, q_r, q_\theta$ and increments in $\nu, \mu_2, \mu_3, \psi$, which are corresponding to axial and polar perturbations, respectively. Here we shall consider the axial type ones. The perturbation equation reads
\begin{eqnarray}
r^4\frac{\partial}{\partial r}\Big(\frac{f(r)}{r^2}\frac{\partial Q}{\partial r}\Big)+\sin^3\theta \frac{\partial}{\partial \theta}\Big(\frac{1}{\sin^3\theta}\frac{\partial Q}{\partial \theta}\Big)-\frac{r^2}{f(r)}\frac{\partial^2 Q}{\partial t^2}=0,
\end{eqnarray}
where
\begin{eqnarray}
Q(t,r,\theta)=e^{i\omega t}Q(r,\theta),\;Q(r,\theta)=r^2f(r)\sin^3\theta Q_{r\theta},\;\; Q_{r\theta}=q_{r,\theta}-q_{\theta,r}.
\end{eqnarray}
Further with $Q(r,\theta)=r\Psi(r)C^{-2/3}_{l+2}$, it can be reduced to Schrodinger wave-like equations:
\begin{eqnarray}\label{schrodinger}
\frac{d^2\Psi}{dr_*^2}+[\omega^2-V(r)]\Psi=0,\;\;dr_*=f(r)dr,
\end{eqnarray}
for gravitational field $\Psi$.
The effective potentials take the form as:
\begin{eqnarray}\label{potential}
V=f(r)\left[\frac{(l+2)(l-1)+2f(r)}{r^2}-\frac{1}{r}\frac{df(r)}{dr}\right].
\end{eqnarray}
 The effective potentials $V$ depend on the value $r$, angular quantum number (multipole momentum) $l$ and the aether coefficient $c_{13}$.

\begin{figure}[ht]
\begin{center}
\includegraphics[width=5.0cm]{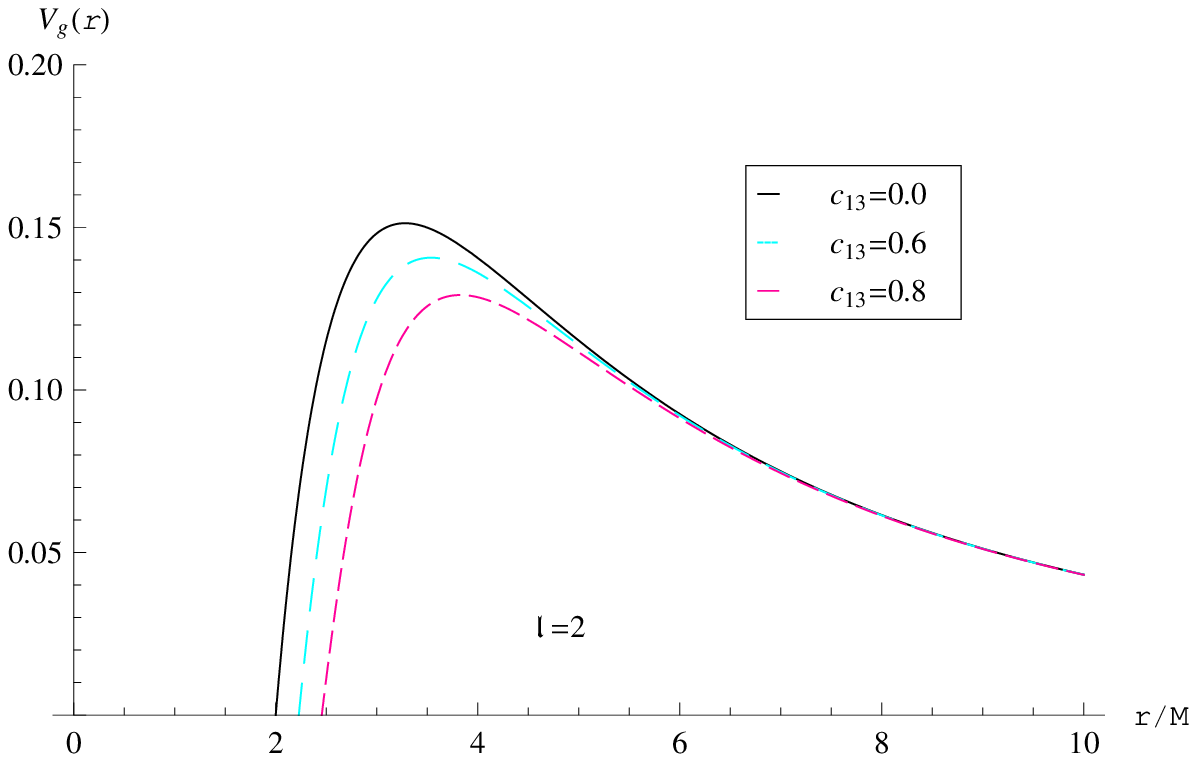}\;\;
\includegraphics[width=5.0cm]{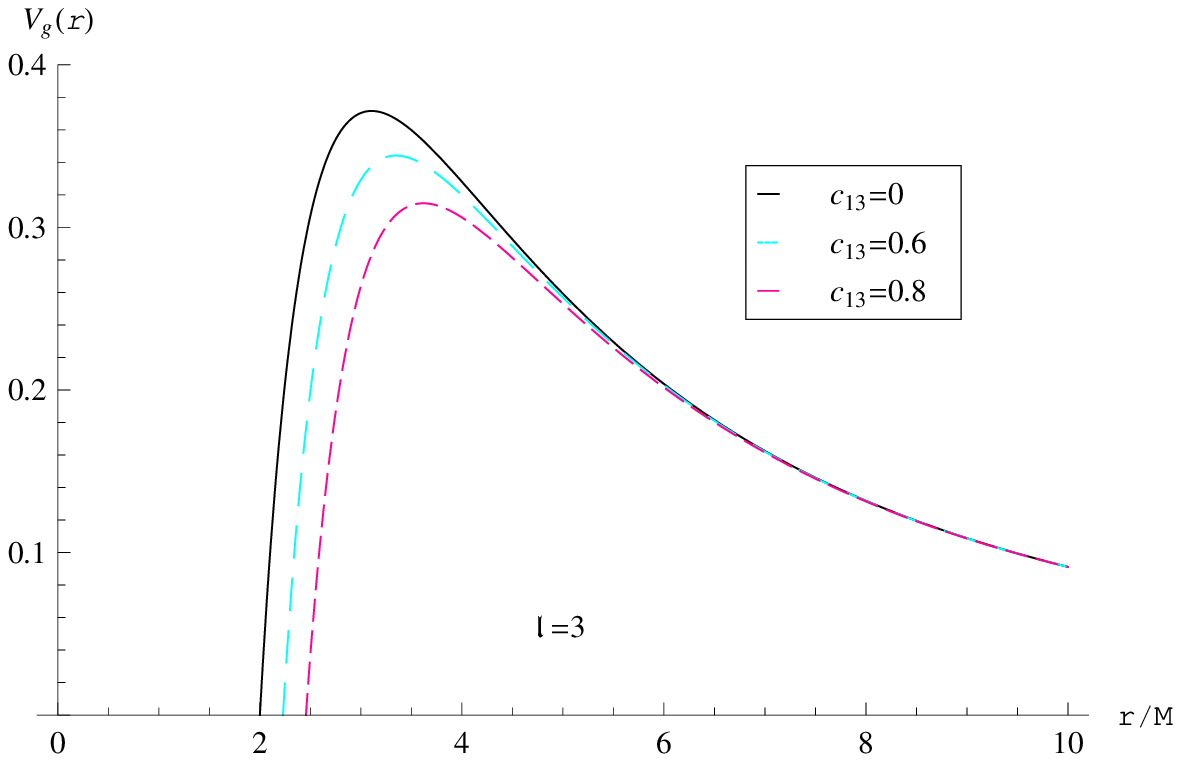}
\;\;\includegraphics[width=5.0cm]{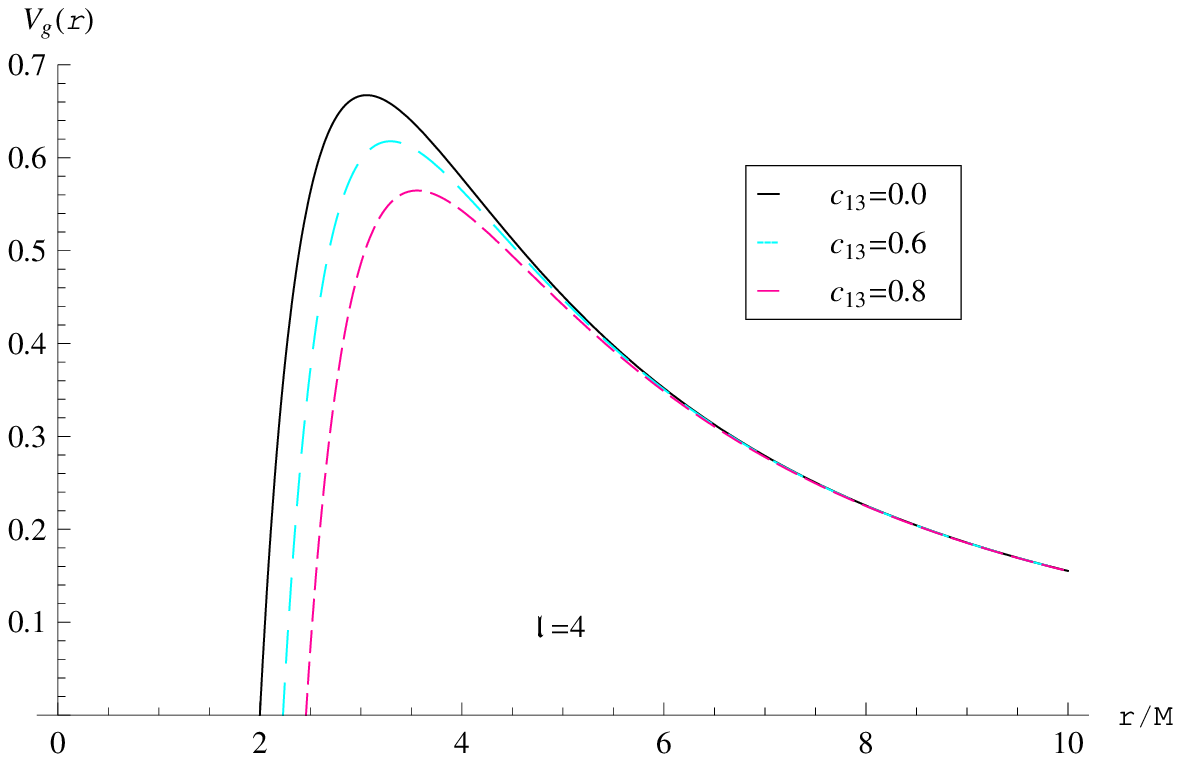}
\caption{The figures are the effective potentials of gravitational field perturbation $V_g$ near the first kind aether black hole $(M=1)$ with different angular quantum number $l$ and coefficients $c_{13}$.}\label{fp1}
 \end{center}
 \end{figure}

From the potential formula (\ref{potential}) and the metric function (\ref{sol1}), the effective potential for the first kind aether black hole is
\begin{eqnarray}
V=\big(1-\frac{2M}{r}\big)\left[\frac{l(l+1)}{r^2}-\frac{6M}{r^3}\right]
+\frac{16M^4I}{r^6}\left[\frac{18M}{r}+96I\frac{M^4}{r^4}-l(l+1)-6\right],
\end{eqnarray}
where the first two terms are Schwarzschild potential, the rests are the aether modified terms, shown in Fig. \ref{fp1}.

In Fig. \ref{fp1},  it is the effective potential of gravitational  field perturbations near the first kind aether black hole. Obviously, if $c_{13}=0$, it can be reduced to those of the Schwarzschild black hole. The peak value gets lower and the turning point shifts to right with $c_{13}$ increasing.
This potential behavior is similar to that of some other Lorentz violating theories.
In the theory of coupling to Einstein's tensor of Reissner-Norstr\"{o}m black hole (see Appendix A), the coupling parameter $\eta$ decreases the peak value of scalar field potential for all $l>0$ \cite{chen201010}. In the massive gravity theory (see Appendix B), the scalar charge $\hat{S}$ decreases the peak value of scalar field potential and shifts its turning point to right \cite{fernando2014}. In the Einstein-Born-Infeld theory (see Appendix D), the Born-Infeld scale parameter $b$ decreases the peak value of scalar and gravitational field potential for all $l$ \cite{fernando}. In the QED-extension limit of SME theory (see Appendix in Ref. \cite{ding2017}), Lorentz violation vector $b_\mu$ also decreases the peak value of Dirac field potential and shifts its turning point to right \cite{chen2006}.
 These properties of the potential will imply that the quasinormal modes possess some different behavior from those of GR case and, some similarities between these Lorentz violating theories.

\begin{figure}[ht]
\begin{center}
\includegraphics[width=5.0cm]{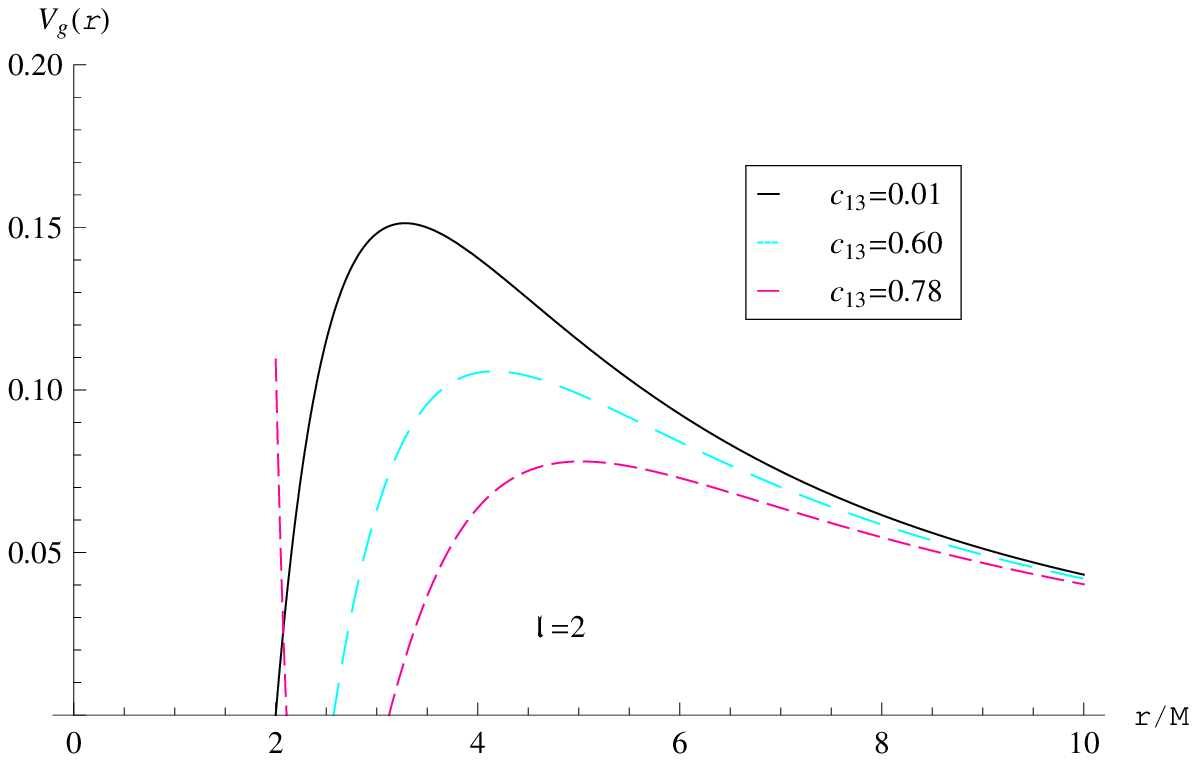}\;\;
\includegraphics[width=5.0cm]{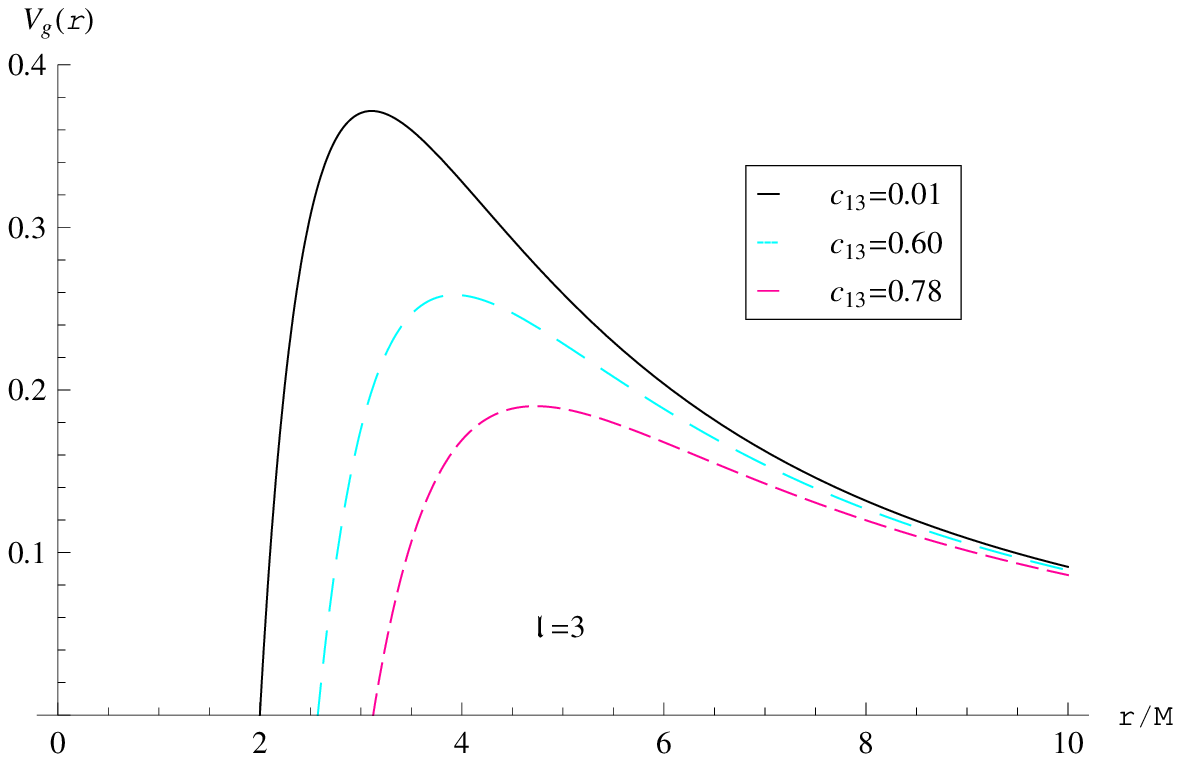}\;\;
\includegraphics[width=5.0cm]{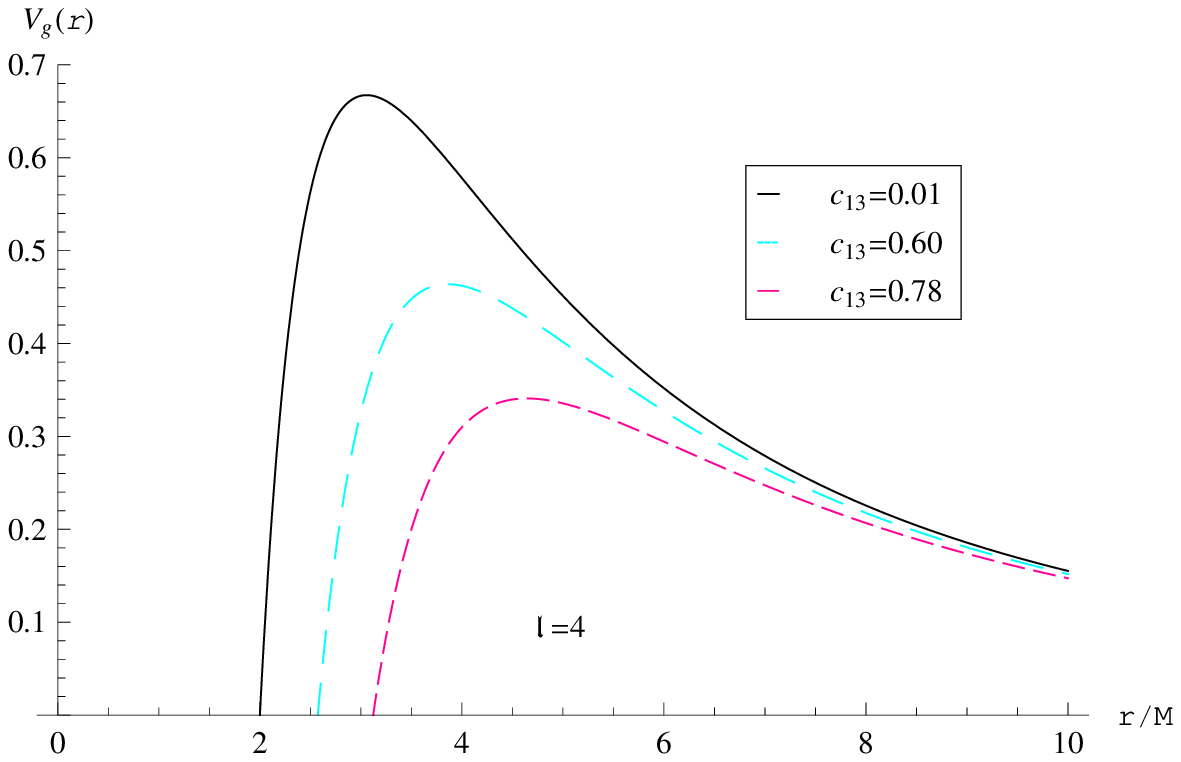}\caption{The figures are the effective potentials of gravitational field perturbation $V_g$ near the second kind aether black hole $(M=1)$ with different coefficients $c_{13}$ and fixed coefficient $c_{14}=0.02$. }\label{fp12}
 \end{center}
 \end{figure}

From the potential formula (\ref{potential}) and the metric function (\ref{sol2}), the effective potential for the second kind aether black hole is
\begin{eqnarray}
V=\big(1-\frac{2M}{r}\big)\left[\frac{l(l+1)}{r^2}-\frac{6M}{r^3}\right]
+\frac{M^2J}{r^4}\left[\frac{14M}{r}+4J\frac{M^2}{r^2}-l(l+1)-4\right],
\end{eqnarray}
where the the first two terms are Schwarzschild potential, the rests are the aether modified terms, shown in Fig. \ref{fp12}.

In Fig. \ref{fp12},  it is the effective potential of the gravitational field perturbation near the second kind aether black hole. It is easy to see that for all $l$, the peak value of the potential barrier gets lower with $c_{13}$ increasing just like the first kind aether black hole and decreasing more quickly. Although both kinds of black holes have  similar potential with coupling constant $c_{13}$, their QNM behaviors with $c_{13}$ will be different with each other partially.

The Schr\"{o}dinger-like wave equation (\ref{schrodinger}) with the effective potential
(\ref{potential}) containing the lapse function $f(r)$ related to the Einstein-aether black holes is not solvable analytically. Since then I now use the sixth-order WKB approximation method to evaluate the quasinormal modes of gravitational field perturbation to the first and second kind aether black holes. The third-order WKB semianalytic method has been proved to be accurate up to around one percent for the real and the imaginary parts of the quasinormal frequencies for low-lying modes with $n<l$ \cite{iyer1,iyer2}. The sixth-order WKB method shows more accurate than third-order one \cite{konoplya2003}. Due to its considerable accuracy for lower lying modes, this method has been used extensively in evaluating quasinormal frequencies of various black holes. In the sixth approximation, the formula for the complex quasinormal frequencies is given by
 \begin{eqnarray}
\omega^2=V_0+\Lambda_2
-i\sqrt{-2V_0''}(\alpha+\Lambda_3+\Lambda_4+\Lambda_5+\Lambda_6),
\end{eqnarray}
where
 \begin{eqnarray}
&&\Lambda_2=\frac{1}{8}\Big(\frac{V_0^{(4)}}{V_0''}\Big)
\Big(\frac{1}{4}+\alpha^2\Big)-\frac{1}{288}\Big(\frac{V_0^{(3)}}{V_0''}\Big)^2(7+60\alpha^2),
\nonumber\\
&&\Lambda_3=\frac{\alpha}{-2V_0''}\Big[\frac{5}{6912}\Big(\frac{V_0^{(3)}}{V_0''}\Big)^4
\Big(77+188\alpha^2\Big)-\frac{1}{384}\Big(\frac{V_0'''^2V_0^{(4)}}{V_0''^3}\Big)(51+100\alpha^2)
\nonumber\\
&&+\frac{1}{2304}\Big(\frac{V_0^{(4)}}{V_0''}\Big)^2(67+68\alpha^2)
\frac{1}{288}\Big(\frac{V_0'''V_0^{(5)}}{V_0''^2}\Big)(19+28\alpha^2)\nonumber\\
&&-\frac{1}{288}\Big(\frac{V_0^{(6)}}{V_0''}\Big)(5+4\alpha^2)\Big],
\end{eqnarray}
and
 \begin{eqnarray}
\alpha=n+\frac{1}{2},\;\;V_0^{(m)}=\frac{d^mV}{dr_*^m}\Big|_{r_*(r_p)},
\end{eqnarray}
the constants $\Lambda_4,\;\Lambda_5,\;\Lambda_6$ are from Ref. \cite{konoplya2003}\footnotemark\footnotetext{The relation between $Q_i$ there and here $V$ is $Q_i=-V_0^{(i)}$.}, $n$ is overtone number and $r_p$ is the turning point value of polar coordinate $r$ at which  the effective potential (\ref{potential}) reaches its maximum. Substituting the effective potential $V$ (\ref{potential}) into the formula above, we can obtain the quasinormal frequencies for the gravitational field perturbations to Einstein-aether black holes. In the next sections, we obtain the quasinormal modes for both kinds of Einstein-aether black holes and analyze their properties.

\section{Quasinormal modes for the first kind aether black hole}

\begin{figure}[ht]
\begin{center}
 \includegraphics[width=7cm]{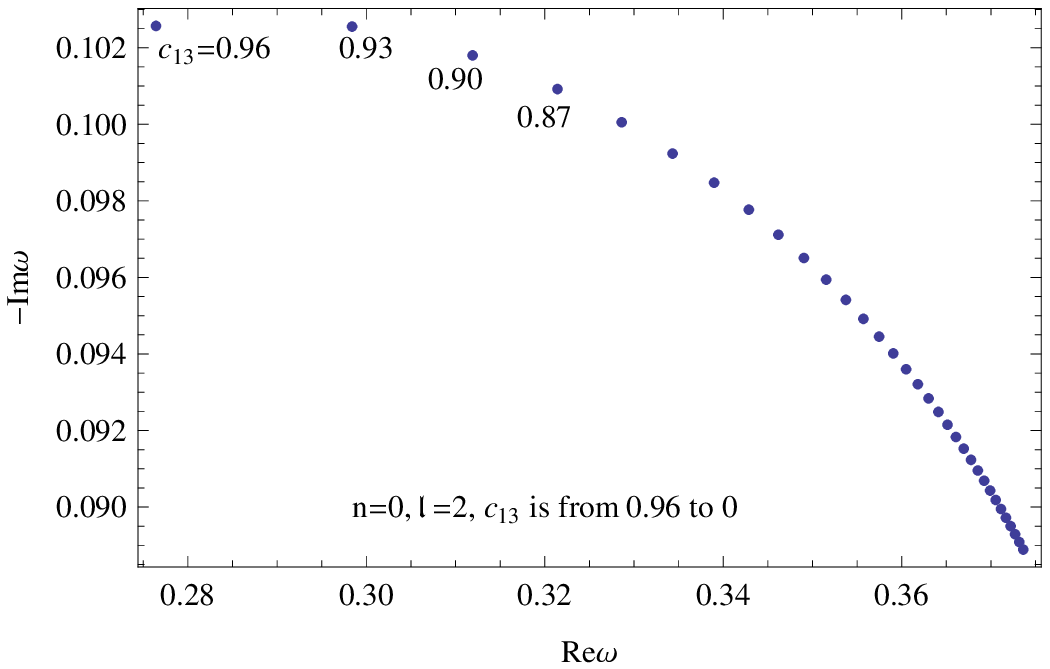}\;\;\;\;\includegraphics[width=7cm]{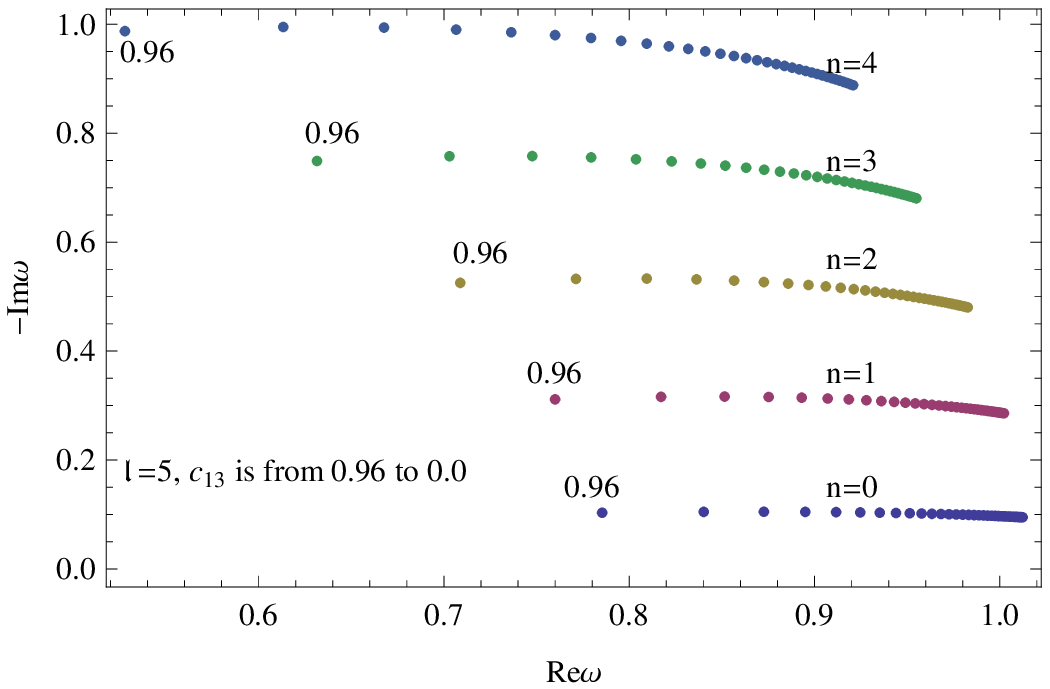}
\caption{The relationship between the real and imaginary parts of quasinormal frequencies of the gravitational field in the background of the first kind aether black hole with the decreasing of $c_{13}$.}\label{fp2}
 \end{center}
 \end{figure}

\begin{figure}[ht]
\begin{center}
 \includegraphics[width=7cm]{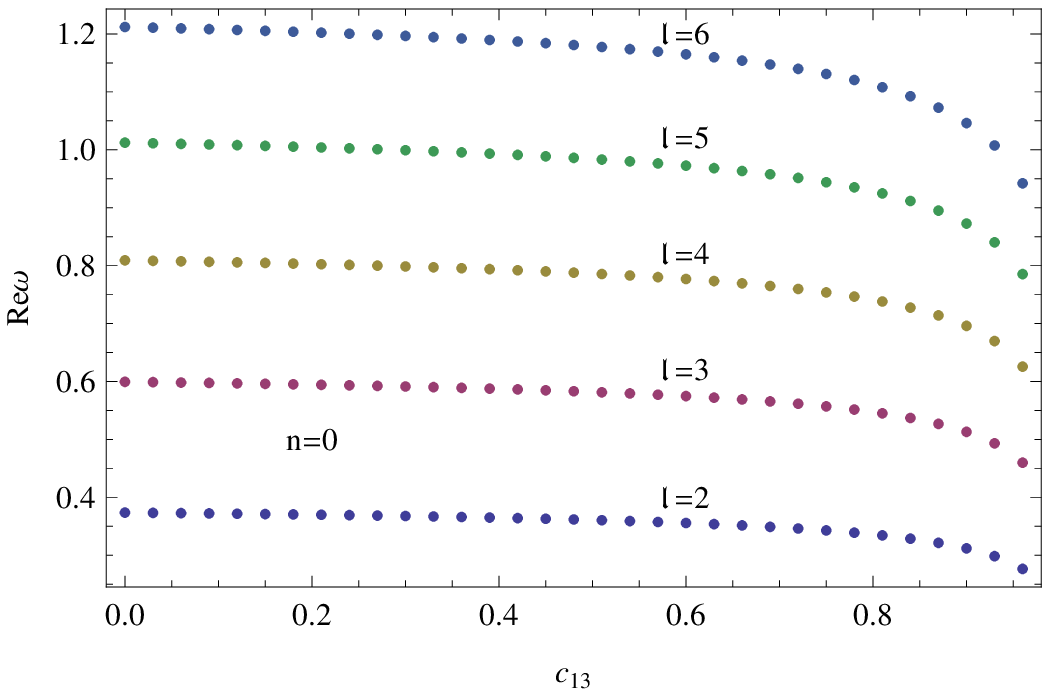}\;\;\;\;\includegraphics[width=7cm]{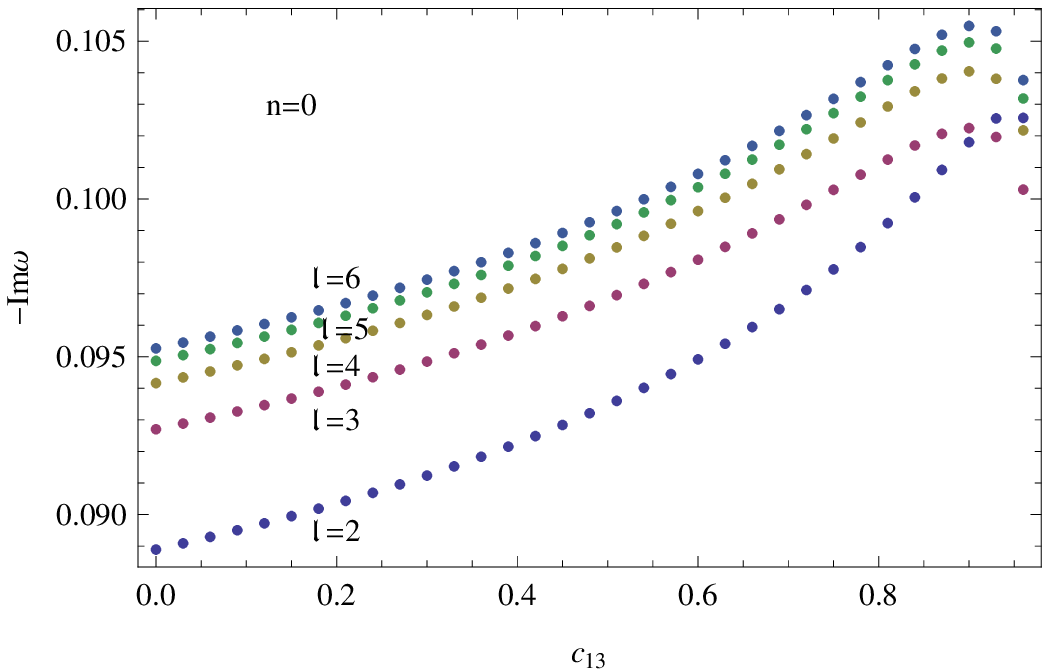}
\caption{The real (left) and imaginary (right) parts of quasinormal frequencies of the gravitational field in the background of the first kind aether black hole with different $c_{13}$.}\label{fp22}
 \end{center}
 \end{figure}

\begin{figure}[ht]
\begin{center}
 \includegraphics[width=7cm]{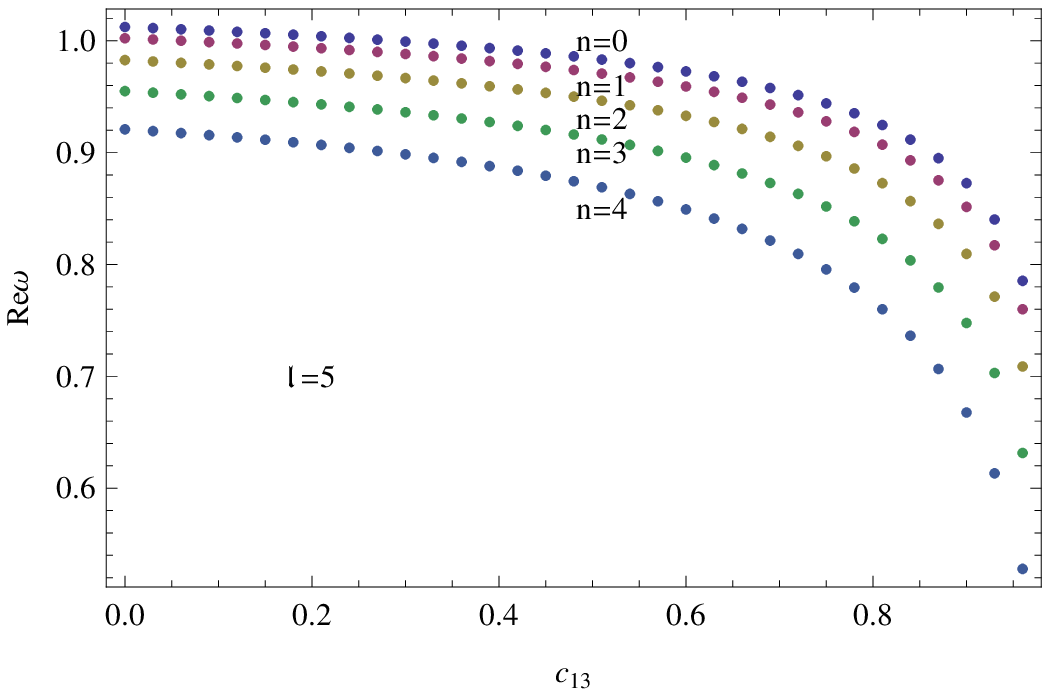}\;\;\;\;\includegraphics[width=7cm]{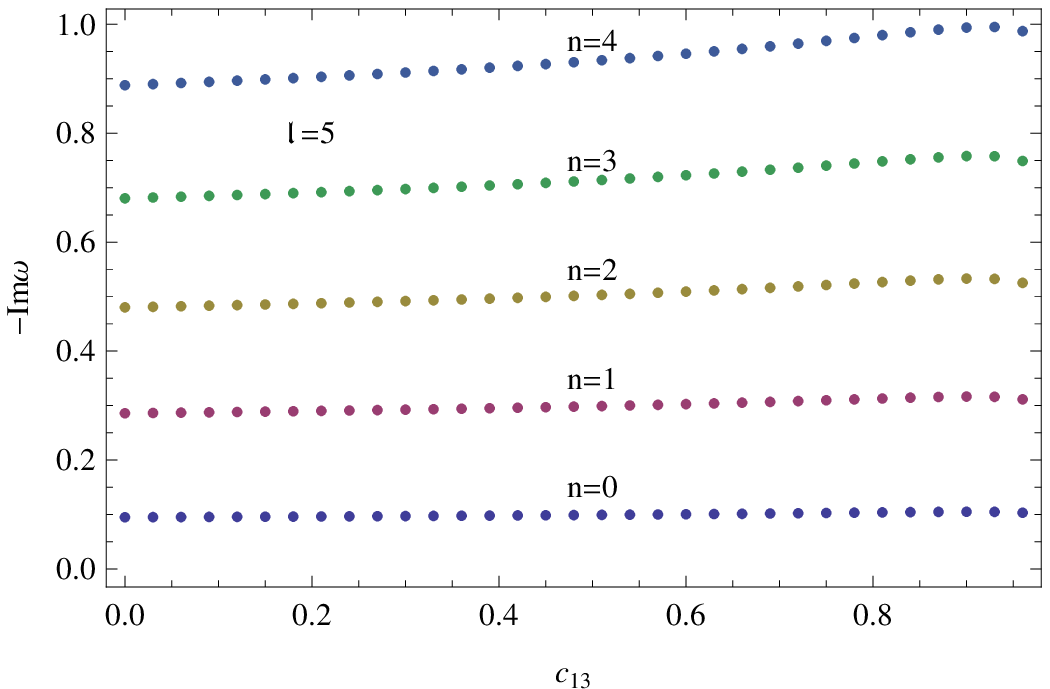}
\caption{The real (left) and imaginary (right) parts of quasinormal frequencies of the gravitational field in the background of the first kind aether black hole with different $c_{13}$.}\label{fp23}
 \end{center}
 \end{figure}

In this section, I study the gravitational field perturbations to the first kind Einstein aether black hole. The perturbation frequencies are shown in Tab. \ref{tab1} and Figs. from \ref{fp2} to \ref{fp23}.
 \begin{table}
 \caption{The lowest overtone $(n=0)$ quasinormal frequencies of the gravitational field in the first kind aether black hole spacetime.}\label{tab1}
 \begin{center}
\begin{tabular}{cccccc}
\hline \hline $c_{13}$&$M\omega(l=2)$&$M\omega(l=3)$&$M\omega(l=4)$&$M\omega(l=5)$&$M\omega(l=6)$\\
\hline
 0.00&\;0.373619-0.088891$i$&\;0.599443-0.092703$i$&\;0.809178-0.094164$i$&\; 1.012300-0.094871$i$&\;1.212010-0.095266$i$\\
  0.15&\;0.371123-0.089949$i$&\;0.595985-0.093676$i$&\;0.804653-0.095142$i$&\; 1.006720-0.095852$i$&\;1.205390-0.096250$i$\\
0.30&\;0.367765-0.091235$i$&\;0.591354-0.094848$i$&\;0.798599-0.096326$i$&\;
0.999260-0.097042$i$&\;1.196530-0.097444$i$\\
0.45&\;0.363006-0.092838$i$&\;0.584808-0.096284$i$&\;0.790050-0.097785$i$&\;
0.988731-0.098514$i$&\;1.184040-0.098924$i$\\
0.60&\;0.355705-0.094918$i$&\;0.574769-0.098074$i$&\;0.776938-0.099617$i$&\;
0.972582-0.100371$i$&\;1.164880-0.100795$i$\\
0.75&\;0.342863-0.097769$i$&\;0.557015-0.100290$i$&\;0.753718-0.101918$i$&\; 0.943965-0.102720$i$&\;1.130900-0.103173$i$\\
0.90&\;0.311907-0.101800$i$&\;0.513073-0.102246$i$&\;0.696009-0.104045$i$&\; 0.872663-0.104961$i$&\;1.046130-0.105484$i$\\
\hline\hline
\end{tabular}
\end{center}
\end{table}

Behaviors with fixed $l,n$ and $c_{13}$. Tab. \ref{tab1} shows the real and the absolute imaginary parts of frequencies are both lower than the electromagnetical and scalar fields perturbations obtained in Ref. \cite{ding2017}, i.e., gravitational$<$electromagnetical$<$scalar. It shows that gravitational perturbations have lowest damping rates and smallest oscillation frequencies.

Behaviors with different $l$. Tab. \ref{tab1} shows that, for fixed $c_{13}$, both the real part and the absolute imaginary part of frequencies increase with the angular quantum number $l$. For large $l$, the imaginary parts approach a fixed value.  These properties are similar to the usual black holes and, also shown in Fig. \ref{fp22}. These increasing behaviors are similar to the electromagnetical field perturbations, but different from scalar field ones whose absolute imaginary part decrease with $l$.
Tab \ref{tab1} also shows the derivations from Schwarzschild black hole. For $l=2$, the decrease in Re$\omega$ is about from $0.7$ percent to $17$ percents, while the increase in $-$Im$\omega$ is about from 1 percent to 15 percents, and may be detected by new generation of gravitational antennas. It will help us to seek LV information in nature in low energy scale.

Behaviors with different overtone numbers $n$. Fig. \ref{fp2} and Fig. \ref{fp23} show that the real parts decrease and the absolute imaginary ones increase with $n$, which are the same as those of the scalar and electromagnetic perturbations \cite{ding2017}.

Behaviors with different aether constants $c_{13}$. For the fixed angular number $l$, or overtone number $n$ with different $c_{13}$ (small $c_{13}$), Tab. \ref{tab1}, Fig. \ref{fp2} and Fig. \ref{fp23} show that the real parts of frequencies decrease, and the absolute imaginary ones increase with the small $c_{13}$ firstly to $c_{13}=0.9$ and then decrease on the contrary, which are similar to the scalar and electromagnetic perturbations \cite{ding2017}. It is different from that of the non-reduced aether black hole \cite{konoplya}, where both all increase with $c_1$. By comparing to Reissner-Norstr\"{o}m black hole, Fig. \ref{fp2} and \ref{fp22} show us a similar behavior that the absolute imaginary part of frequencies increases for small parameter $c_{13}$ or $Q$, and then decrease in the region of large parameter \cite{konoplya2002}. The only difference is that the real part decreases here for all $c_{13}$ and increases there for all $Q$.

By comparing to some other LV models --- nonminimal  coupling theory to Eistein's tensor, massive gravity theory and the QED-extension limit of SME, the above gravitational field QNMs properties with $c_{13}$ are similar to those of them: the scalar field QNMs with coupling constant $\eta$ \cite{chen201010}, the scalar field QNMs with the scalar charge $\hat{S}$ \cite{fernando2014} and Dirac field QNMs with LV coefficient $b_\mu$ \cite{chen2006}, respectively. These similarities are that the real part decreases while the absolute imaginary part increases with the given LV coefficient $\eta,\hat{S}$ and $b$.

In the theory of QED-extension limit of SME, local LV coefficient $b_\mu$ is introduced in a matter sector, while in Einstein-aether theory, the nonminimal coupling theory and massive gravity, local LV is in a gravity sector. For the former, LV matter perturbs to LI black hole --- Schwarzschild black hole and produces QNMs. For the latter, LI matter perturbs to LV black hole --- Einstein-aether black hole, the modified Reissner-Norstr\"{o}m black hole and massive gravity black hole, and then produces QNMs. These similarities between different backgrounds may imply some common property of LV coefficient on QNMs, i.e., in presence of LV, the perturbation field oscillation damps more rapidly, and its period becomes longer. They also motivate us to the further theoretical  study on the possible intrinsic connections between them.

\section{Quasinormal modes for the second kind aether black hole}

\begin{figure}[ht]
\begin{center}
\includegraphics[width=7.0cm]{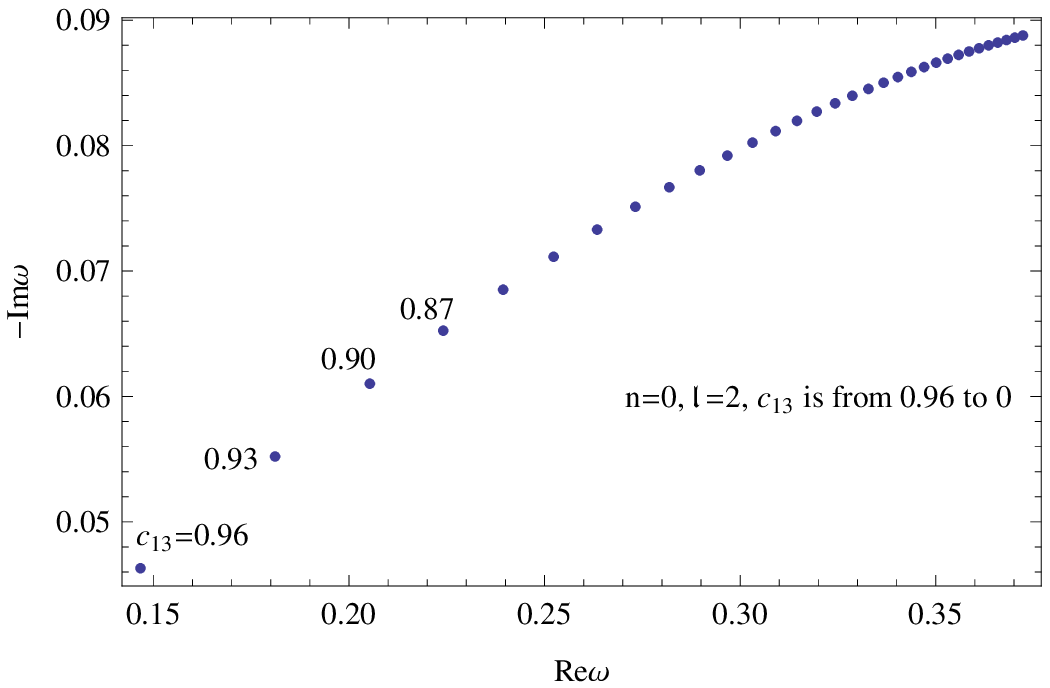}\;\;\;\;
 \includegraphics[width=7.0cm]{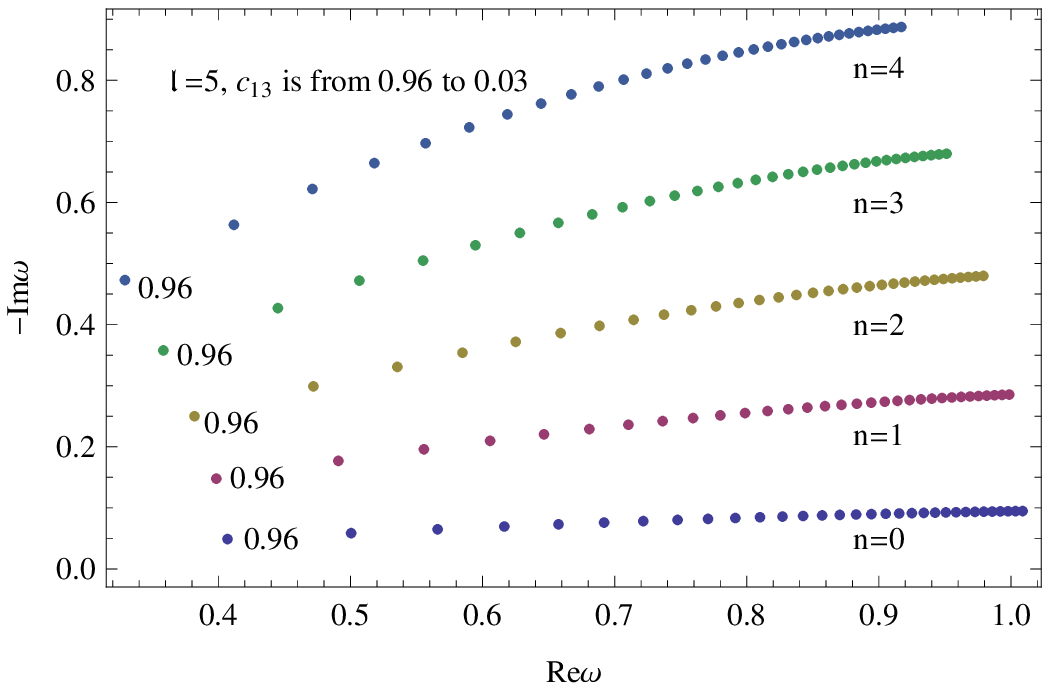}
\caption{The relationship between the real and imaginary parts of quasinormal frequencies of the gravitational field in the background of the second kind aether black hole with the decreasing of $c_{13}$. }\label{fp3}
 \end{center}
 \end{figure}

\begin{figure}[ht]
\begin{center}
\includegraphics[width=7.0cm]{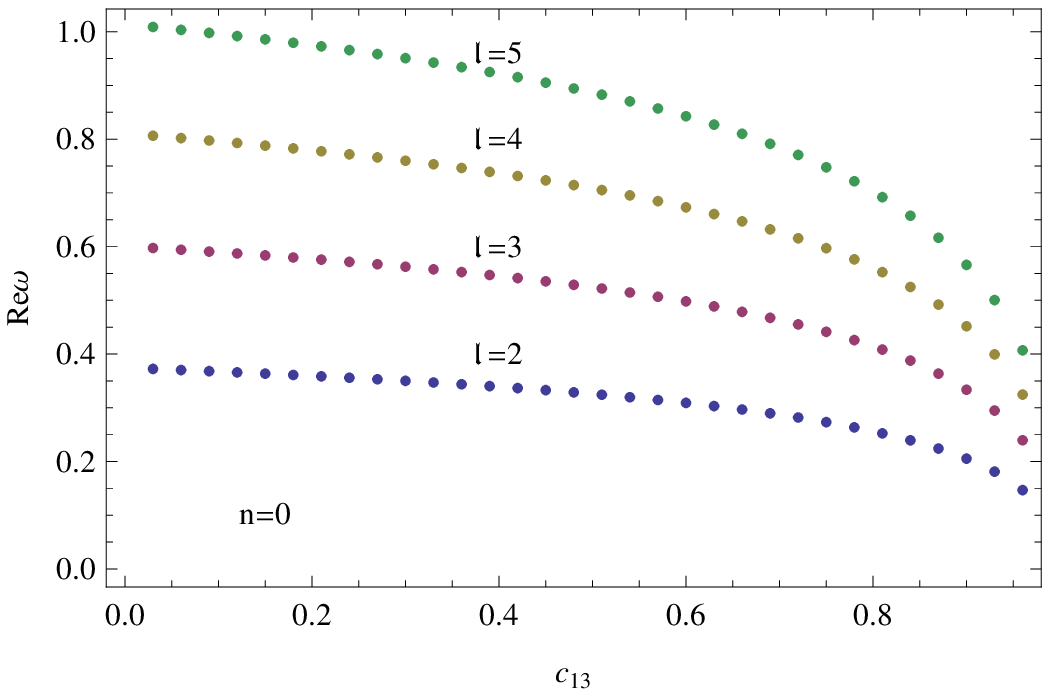}\;\;\;\;
 \includegraphics[width=7.0cm]{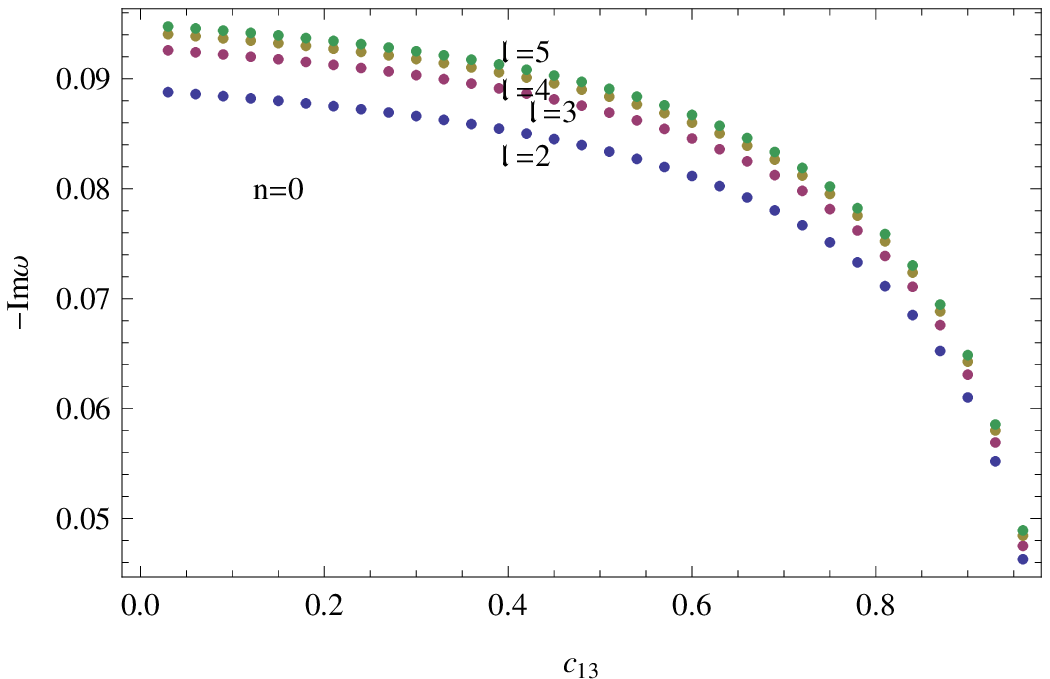}
\caption{The real (left) and imaginary (right) parts of quasinormal frequencies of the gravitational field in the background of the second kind aether black hole with different $c_{13}$. }\label{fp32}
 \end{center}
 \end{figure}

\begin{figure}[ht]
\begin{center}
\includegraphics[width=7.0cm]{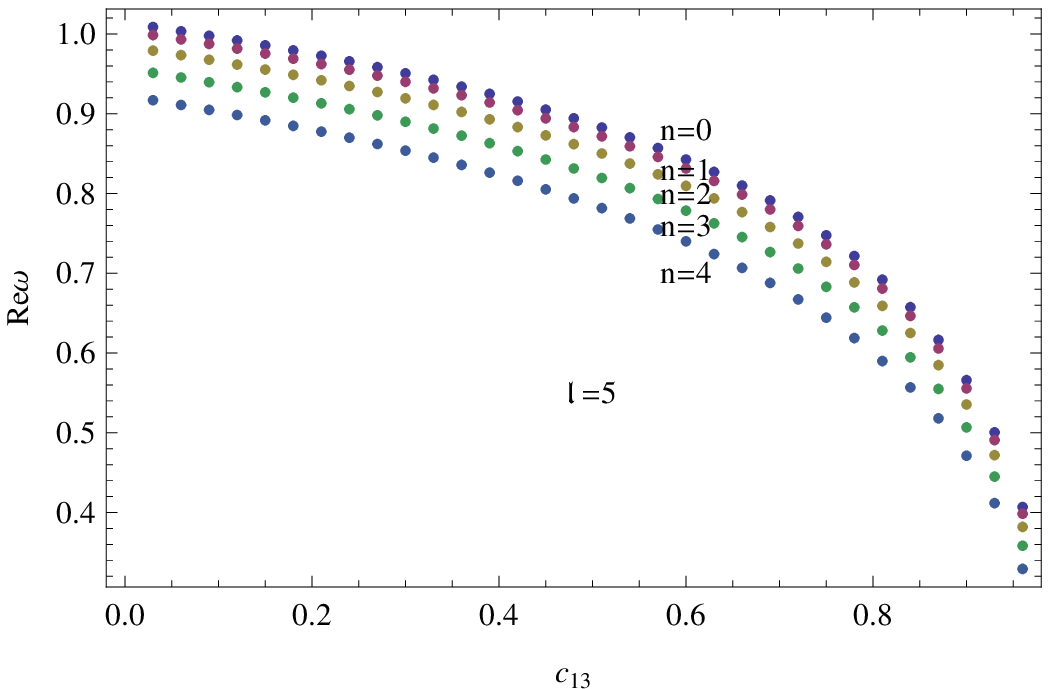}\;\;\;\;\;
 \includegraphics[width=7.0cm]{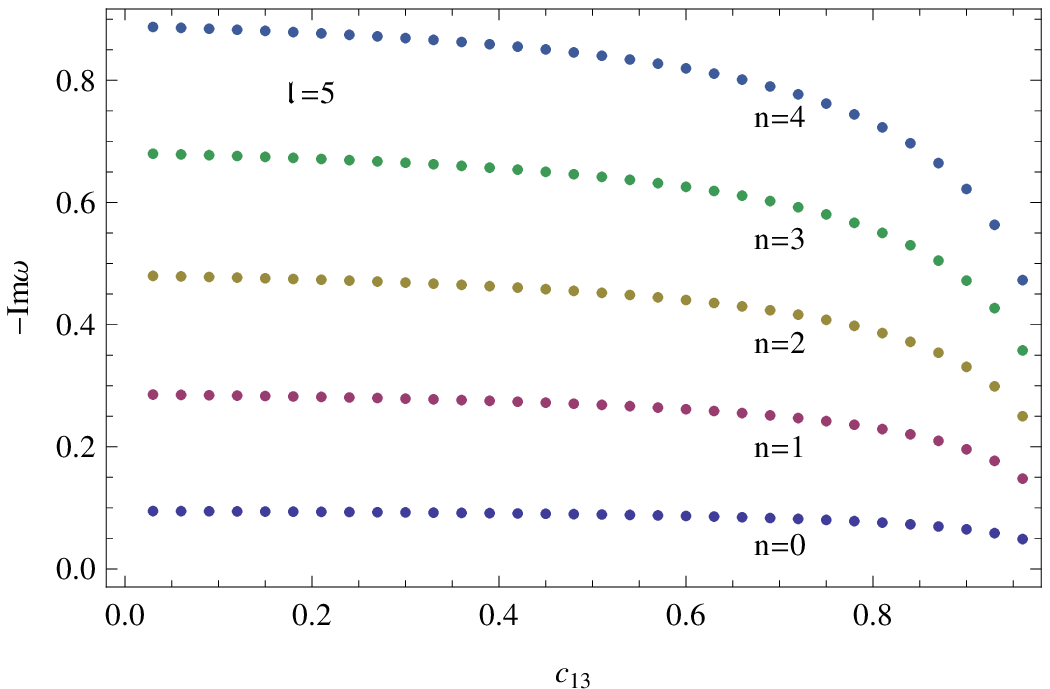}
\caption{The relationship between the real and imaginary parts of quasinormal frequencies of the gravitational field in the background of the second kind aether black hole with the decreasing of $c_{13}$ } \label{fp33}
 \end{center}
 \end{figure}

In this section, I study the gravitational field perturbations to the second kind of Einstein-aether black hole with fixed $c_{14}=0.02$.  Their behaviors are shown in Tab. \ref{tab2} and Figs. from \ref{fp3} to \ref{fp33}.
 \begin{table}
 \caption{The lowest overtone ($n=0$) quasinormal frequencies of the gravitational field in the second kind aether black hole spacetime with fixed $c_{14}=0.02$.}\label{tab2}
 \begin{center}
\begin{tabular}{cccccc}
\hline \hline $c_{13}$ &$M\omega(l=2)$ &$M\omega(l=3)$&$M\omega(l=4)$ &$M\omega(l=5)$&$M\omega(l=6)$ \\
\hline
0.01&\;0.373619-0.088891$i$&\;0.599443-0.092703$i$&\;0.809178-0.094164$i$&\; 1.012300-0.094871$i$&\;1.212010-0.095266$i$\\
  0.15&\;0.363508-0.087996$i$&\;0.583585-0.091775$i$&\;0.787938-0.093240$i$&\; 0.985828-0.093950$i$&\;1.180390-0.094348$i$\\
0.30&\;0.350114-0.086611$i$&\;0.562563-0.090330$i$&\;0.759762-0.091795$i$&\;
0.950709-0.092508$i$&\;1.138430-0.092909$i$\\
0.45&\;0.332795-0.084518$i$&\;0.535354-0.088129$i$&\;0.723264-0.089587$i$&\;
0.905196-0.090301$i$&\;1.084040-0.090703$i$\\
0.60&\;0.309073-0.081156$i$&\;0.498024-0.084570$i$&\;0.673136-0.086006$i$&\;
0.842654-0.086713$i$&\;1.009280-0.087113$i$\\
0.75&\;0.273208-0.075123$i$&\;0.441421-0.078143$i$&\;0.597022-0.079520$i$&\; 0.747622-0.080205$i$&\;0.895630-0.080593$i$\\
0.90&\;0.205326-0.061015$i$&\;0.333580-0.063088$i$&\;0.451685-0.064270$i$&\; 0.565957-0.064866$i$&\;0.678232-0.065208$i$\\
\hline\hline
\end{tabular}
\end{center}
\end{table}

The $l$-dependant behaviors. Tab. \ref{tab2} shows that, for fixed $c_{13}$, both the real and the absolute imaginary parts of frequencies increase with the angular quantum number $l$. For large $l$, the imaginary parts approach a fixed value which is similar to the first kind aether black hole and,  also shown in Fig. \ref{fp32}.
Tab \ref{tab2} also shows the derivations from Schwarzschild black hole. For $l=2$, the decrease in Re$\omega$ is about from $3$ percents to $45$ percents, while the decrease in $-$Im$\omega$ is about from 1 percent to 31 percents, both are bigger than the first kind aether black hole, and may be detected by new generation of gravitational antennas.

The $n$-dependant behaviors. For different overtone numbers $n$, Fig. \ref{fp3} and Fig. \ref{fp33} show that the real parts decrease and the absolute imaginary ones increase with $n$, which is the same as that of the first kind aether black hole.

The $c_{13}$-dependant behaviors. For the fixed angular number $l$, or overtone number $n$,  Tab. \ref{tab2}, Fig. \ref{fp3} and Fig. \ref{fp32} show that both the real and  the absolute imaginary parts of frequencies all decrease with $c_{13}$ increasing, which are completely different from that of the non-reduced aether black hole \cite{konoplya}, or partially different from those of the first kind aether black hole. And they are the same as those of the scalar and electromagnetic perturbations obtained in Ref. \cite{ding2017}. This property of both decreases is similar to that of the noncommutative Schwarzschild black hole \cite{liang} and Einstein-Born-Infeld black hole \cite{fernando}.

It is interesting that the three kinds of black holes' masses/charge are all modified. For the second kind aether black hole, the aether field contributes the spacetime mass as $M_{\ae}=-c_{14}M_{ADM}/2$ \cite{ding}, where $M_{ADM}$ is its Komar mass. For the noncommutative Schwarzschild black hole, the pointlike mass is replaced by smeared mass distribution so that there is no singularity at the origin $r=0$. For the Einstein-Born-Infeld black hole, the pointlike charge is replaced by non-linear distribution also leads to no singularity the electromagnetic field at the origin. Is Einstein-Born-Infeld theory also Lorentz breaking? It is an open issue that needs to be studied in the future.

 Noncommutative theory has a strong quantum gravity motivation.
The similarities of QNMs between it and the second kind aether black hole show us that Lorentz symmetry should be given up in a potential quantum gravity.

\section{Summary}

The local Lorentz violation in the gravity sector should show itself in radiative processes around black holes.  And the similar behaviors of QNMs with the corresponding parameters in some LV theories can lead us to further understand them.

In this paper, I study on QNMs of the gravitational field perturbations to Einstein-aether black holes.  In Einstein-aether theory, Lorentz symmetry is broken by the existence of the aether field $u^a$. This aether field doesn't affect the spacetime mass of the first kind aether black hole, but contributes mass as $M_{\ae}=-c_{14}M_{ADM}/2$ in the second kind aether black hole spacetime.

To the effective potential, when $c_{13}$ increases, the turning points for both kinds of black holes always become larger, and the value of their peak becomes lower. However, their QNMs are different from each other that show their complexity. For the first kind aether black hole, the real part of gravitational QNMs becomes smaller with all $c_{13}$ increase. The absolute value of imaginary part of QNMs becomes bigger with small $c_{13}$ increase, and then decreases with big $c_{13}$. For the second kind aether black hole, both decrease with $c_{13}$.  For the non-reduced aether hole \cite{konoplya}, both real part and the absolute value of imaginary part of QNMs increase with $c_1$.

Firstly by comparing to Schwarzschild black hole, the first kind aether black holes have larger damping rate and the second ones have lower damping rate. They all have smaller real oscillation frequency of QNMs.  If the breaking of Lorentz symmetry is not very small, the derivation of QNMs from Schwarzschild values might be observed in the near future gravitational wave events and, detected by new generation of gravitational antennas.

Secondly, by comparing to some other LV gravity theories, the properties of QNM behaviors for the first kind Einstein-aether black hole with $c_{13}$ are similar to the behaviors of the scalar field QNMs with coupling constant $\eta$ \cite{chen201010}, the scalar field QNMs with the scalar charge $\hat{S}$ \cite{fernando2014} and Dirac field QNMs with LV coefficient $b$ \cite{chen2006}. These similarities between different backgrounds may imply some connections between  Einstein-aether theory, the non-minimal coupling theory, massive gravity theory and the QED-extension limit of SME, i.e., LV in gravity sector and LV in matter sector will make quasinormal ringing of black holes damping more rapidly and its period becoming longer.

The properties of QNM behaviors for the second Einstein-aether black hole with $c_{13}$ are similar to the scalar and gravitational field QNMs of Einstein-Born-Infeld black holes with the Born-Infeld parameter $b$ \cite{fernando} and, the scalar, gravitational, electromagnetic and Dirac fields QNMs  of noncommutative Schwarzschild black hole \cite{liang} with the spacetime noncommutative parameter $\vartheta$. In this case, LV in gravity sector affects spacetime mass and will make quasinormal ringing of black holes damping more slowly and its period becoming longer.
 Noncommutative theory has a strong quantum gravity motivation.
The similarities of QNMs between it and the second kind aether black hole show us that Lorentz symmetry should be given up in a model merging SM and GR.

Thirdly by comparing to Ref. \cite{ding2017}, the real and the absolute imaginary parts of QNMs are lower than the electromagnetical and scalar fields perturbations, i.e., gravitational$<$electromagnetical$<$scalar.
For fixed $c_{13}$, like the electromagnetical field, both the real part and the absolute imaginary part of frequencies increase with $l$.

\begin{acknowledgments}
The author would like to thank Emanuele Berti, Jiliang Jing, Songbai Chen, R. A. Konoplya and Kai Lin  for enthusiastic helps.  This work was supported by the National Natural Science Foundation
of China (grant No. 11247013), Hunan Provincial Natural Science Foundation of China (grant No. 2015JJ2085), and the fund under grant No. QSQC1708.
\end{acknowledgments}

\appendix
\section{Non-minimal coupling theory}
Scalar fields are relatively simple, which allows us to probe the detailed features of the more complicated physical system. The non-minimal coupling between scalar field and higher order terms in the curvature naturally rise to improve the early inflationary models and could contribute to solve the dark matter problem. Ding {\it et al} extended this coupling theory to dynamical gravity \cite{ding2010}. Chen {\it et al} found that the coupling theory between the kinetic term of scalar field $\psi$ and the Einstein's tensor $G^{\mu\nu}$
\begin{eqnarray}\
S=\int d^4x\sqrt{-g}\big[\frac{R}{16\pi G}+\frac{1}{2}g^{\mu\nu}\partial_{\mu}\psi\partial_\nu\psi
+\frac{\eta}{2}G^{\mu\nu}\partial_{\mu}\psi\partial_\nu\psi\big]
\end{eqnarray}
 is Lorentz violation \cite{chen2015}, where $\eta$ is a coupling constant. This coupling modifies Reissner-Nordstr\"{o}m spacetime as
 \begin{eqnarray}\
ds^2=\left(1-\eta \frac{Q^2}{r^4}\right)\left[-f(r)dt^2+\frac{1}{f(r)}dr^2\right]+
\left(1+\eta \frac{Q^2}{r^4}\right)r^2(d\theta^2+\sin^2\theta d\phi^2),
\end{eqnarray}
where $f(r)=1-2M/r+Q^2/r^2$.
  In 2010, Chen {\it et al} studied the dynamical evolution of a scalar field coupling to Einstein's tensor in Reissner-Nordstr\"{o}m spacetime \cite{chen201010}.
\section{Massive gravity}
To explain the accelerated expansion of the Universe without dark energy and dark matter components, one can employ spontaneous breaking Lorentz symmetry by condensates of four scalar fields $\phi^0$ and $\phi^i$ coupled to gravity. Then the gravitons acquire a mass, which is similar to the Higgs mechanism \cite{fernando2014}. The action is
\begin{eqnarray}
S=\int d^4x\sqrt{-g}\big[\frac{R}{16\pi}+\Lambda^4\mathcal{F}(X,W^{ij})\big],
\end{eqnarray}
where $R$ is spacetime curvature scalar, $X$ and $W^{ij}$ are defined by the four scalar fields $\phi^0$ and $\phi^i$ as
\begin{eqnarray}
X=\frac{\partial^{\mu}\phi^0\partial_{\mu}\phi^0}{\Lambda^4},\;
W^{ij}=\frac{\partial^{\mu}\phi^i\partial_{\mu}\phi^j}{\Lambda^4}-
\frac{\partial^{\mu}\phi^i\partial_{\mu}\phi^0\partial^{\nu}\phi^j\partial_{\nu}\phi^0}
{\Lambda^8X}.
\end{eqnarray}
Here the constant $\Lambda$ has the dimension of mass; $\phi^0=\Lambda^2[t+h(r)]$ and $\phi^i=\Lambda^2x^i$ which are called Goldstone fields. There is a static spherically symmetric solution in which the metric function is
\begin{eqnarray}
f(r)=1-\frac{2M}{r}-\frac{\hat{S}}{r^{\lambda}},
\end{eqnarray}
where $\hat{S}$ is a scalar charge, $\lambda$ is a positive constant. When the scalar charge $\hat{S}=0$,  Schwarzschild solution is recovered.
And in the limit $\lambda\rightarrow\infty$, it approaches to Schwarzschild behavior in the large $r$ region.
With this metric function, Fernando {\it et al} studied the QNMs of spherically symmetric black hole \cite{fernando2014} in the massive gravity theory.

\section{Noncommutative gravity theory}

In quantization of spacetime to construct quantum gravity theory, a common conjecture is that the algebra of spacetime coordinates is actually noncommutative. The most familiar form of Noncommutative gravity theory is that the spacetime coordinates acquire the commutative relation
\begin{eqnarray}
 [x^{\mu},\;x^{\nu}]=i\vartheta^{\mu\nu},
\end{eqnarray}
where $\vartheta^{\mu\nu}$ is a real, antisymmetric, and constant tensor which determines the fundamental cell discretization of spacetime much in the same way as Planck constant $\hbar$ discretizes the phase space $[x_i,p_j]=i\hbar\delta_{ij}$. Lorentz symmetry is intrinsically broken by virtue of  nonzero $\vartheta^{\mu\nu}$\cite{carroll2001}.
By using $\vartheta^{\mu\nu}=\vartheta $diag$(\epsilon_1, ... ,\epsilon_{D/2})$, Nicolini {\it et al} derived a noncommutative Schwarzschild black hole solution by a metric function
\begin{eqnarray}\
f(r)=1-\frac{4M}{r\sqrt{\pi}}\gamma(3/2,r^2/4\vartheta),
\end{eqnarray}
where $\gamma(r,\vartheta)$ is the lower incomplete Gamma function and $\vartheta$ is a real constant denotes spacetime noncommutativity. It leads to the mass distribution $m(r)=2M\gamma(3/2,r^2/4\vartheta)/\sqrt{\pi}$, where $M$ is the total mass of the source. In another word, it is that the pointlike mass $M$ is replaced by smeared mass $m$ which leads to no singularity at origin $r=0$, i.e., its spacetime curvature scalar
\begin{eqnarray}
R|_{r=0}=\frac{4M}{\sqrt{\pi}\vartheta^{3/2}}.
\end{eqnarray}
 When $\vartheta\rightarrow0$, a Schwarzschild black hole is recovered. With this metric function, Jun Liang studied QNMs of a noncommutative geometry inspired Schwarzschild black hole \cite{liang}.

\section{Einstein-Born-Infeld gravity theory}
Born-Infeld theory plays an important role in string theory: it arises naturally in open superstrings and in D-branes. Einstein-Born-Infeld black hole solutions may also play a role in understanding the black hole in the deformed  Ho\u{r}ava-Lifshitz gravity\cite{myung}. Is Einstein-Born-Infeld theory also Lorentz breaking? It is an open issue that needs to be studied in the future. In Born-Infeld nonlinear electrodynamics theory, the action is
\begin{eqnarray}
S=\int d^4x\sqrt{-g}\left[\frac{R}{16\pi G}+\frac{1}{b}\Big(1-\sqrt{1+2bF^{\mu\nu}F_{\mu\nu}}\Big)\right],
\end{eqnarray}
where $F^{\mu\nu}$ is the electromagnetic field strength.
For the spherically symmetric case, the electric field is
\begin{eqnarray}
E(r)=Q/\sqrt{r^4+\frac{b^2Q^2}{4}},
\end{eqnarray}
where $b$ is a Born-Infeld parameter of dimensions length square. Then it can avoid the singularity at the position of a pointlike charge.
The metric function of static charged black hole solution in Einstein-Born-Infeld theory is
\begin{eqnarray}
f(r)=1-\frac{2M}{r}+\frac{r^2}{6b}-\frac{1}{r}\int\sqrt{\frac{Q^2}{b}+\frac{r^4}{4b^2}}dr,
\end{eqnarray}
which is regular at the origin.
In the limit $b\rightarrow0$\footnotemark\footnotetext{Here $b$ relates to $\beta$ in ref. \cite{fernando} by $b=1/4\beta^2$.}, the Reissner-Nordstr\"{o}m solution is recovered,
\begin{eqnarray}
f(r)=1-\frac{2M}{r}+\frac{Q^2}{r^2}.
\end{eqnarray}
With this metric function, Fernando and Chen {\it et al} studied the QNMs of spherically symmetric Einstein-Born-Infeld black hole\cite{fernando}.

\end{document}